\documentclass[pra,aps,showpacs,amsmath,amssymb,floats,twocolumn]{revtex4}
\usepackage{graphics,dcolumn,bm}
\usepackage{color}
\usepackage[hypertex]{hyperref}

\begin{document}
\title{Optimal Control of the Strong-Field Ionization of Silver Clusters in
  Helium Droplets}

\author{N.~X.~Truong$^1$, P.~Hilse$^2$, S.~G\"ode$^1$, A.~Przystawik$^1$,
  T.~D\"oppner$^1$, Th.~ Fennel$^1$, Th.~Bornath$^1$, J.~Tiggesb\"aumker$^1$,
  M.~Schlanges$^2$, G.~Gerber$^3$, and K.~H.~Meiwes-Broer$^1$}
\affiliation{$^1$ Institut f\"ur Physik, Universit\"at Rostock, 18051 Rostock,
  Germany} \affiliation{$^2$ Institut f\"ur Physik,
  Ernst-Moritz-Arndt-Universit\"at Greifswald, 17489 Greifswald, Germany}
\affiliation{$^3$ Physikalisches Institut, Universit\"at W\"urzburg, 97074
  W\"urzburg, Germany } \date{\today}

\begin{abstract}
  Optimal control techniques combined with femtosecond laser pulse shaping are
  applied to steer and enhance the strong-field induced emission of highly
  charged atomic ions from silver clusters embedded in helium nanodroplets.
  With light fields shaped in amplitude and phase we observe a substantial
  increase of the Ag$^{q+}$ yield for $q>10$ when compared to
  bandwidth-limited and optimally stretched pulses. A remarkably
  simple double-pulse structure, containing a low-intensity prepulse and a
  stronger main pulse, turns out to produce the highest atomic charge states
  up to Ag$^{20+}$. A negative chirp during the main pulse hints at dynamic
  frequency locking to the cluster plasmon. A numerical optimal control study
  on pure silver clusters with a nanoplasma model converges to a similar pulse
  structure and corroborates, that the optimal light field adapts to the
  resonant excitation of cluster surface plasmons for efficient ionization.
\end{abstract}
\email{josef.tiggesbaeumker@uni-rostock.de}
\maketitle
\section{Introduction}
\label{intro}
Atomic clusters in intense laser pulses have received considerable attention
since the mid-90's because of the potential for several promising
applications, such as the controlled generation of highly charged atomic
ions~\cite{SnyPRL96}, energetic electrons~\cite{ShaPRL96},
x-rays~\cite{McPN94}, or even nuclear particles~\cite{DitN99}. Besides
technological interest, the absence of hidden dissipation channels makes
clusters to a valuable testing ground for exploring many-particle effects in
strong-field laser-matter interactions~\cite{SaaJPB06} .

A remarkable feature of clusters exposed to intense near-infrared (NIR)
femtosecond laser pulses is the extremely efficient absorption of radiation
energy due to the coupling of the light field to a dense transient
nanoplasma~\cite{DitPRL97a}. The resulting highly nonlinear cluster response
has been studied theoretically by several groups and it is widely accepted
that resonant collective electron heating by excitation of Mie surface
plasmons is a key mechanism for high absorption as well as strong cluster
ionization~\cite{DitPRA96,MilPRE01,KunPP08,SaaPRL03,MikPRA08}.  However, when
considering NIR laser pulses, the energy of the plasmon resonance in metallic
as well as in preionized rare-gas clusters is typically well above the laser
photon energy of $1.5\,{\rm eV} $ in early stages of the interaction.  For
example, a plasmon energy of about $4.0\,\rm eV$ is found for silver
clusters~\cite{TigPRA93} in the ground state and may be further increased by
laser-induced inner ionization. Noting the density dependence of the
Mie-plasmon energy ($\omega_{\rm mie}\sim\sqrt{\rho_{\rm I}}$), where
$\rho_{\rm I}$ is the charge density of the ionic background), resonant
collective heating requires a certain cluster expansion induced by ionization
and non-resonant heating. Obviously, such sequence of ionization, expansion,
and transient resonance heating may be realized and resolved with very
different pulse shapes, e.g., by excitation with a long pulse or with multiple
well-separated pulses, respectively~\cite{ZwePRA99}. Likewise, the interaction
process can be sensitively tuned by modifying the laser parameters, e.g., by
changing the pulse duration, envelope, or chirp.

Indeed, several experiments have demonstrated that such simplified control
techniques allow the enhancement of certain decay channels by optimizing the
light field. In particular, cluster excitation with (i) stretched pulses
having an optimal duration or with (ii) dual pulses having an optimal delay
substantially support the generation of highly charged, energetic
ions~\cite{KoePRL99,DoePRL05,KumPRA02,HeiLP07} and fast
electrons~\cite{FenPRL07a,KumPRA03}. At optimal pulse length, ion recoil
energies can be further increased by (iii) introducing a weak additional
shoulder in the leading edge of the laser pulse~\cite{MenOE03} or by (iv)
using a negative pulse chirp~\cite{FukPRA03}.

As a next logical step, optimal control techniques based on iterative feedback
with genetic algorithms can be used to further explore and optimize the
cluster excitation process. Considering the strong pulse-structure dependence
obtained in previous studies (i-iv), laser-cluster interactions provide ideal
grounds to test and apply control schemes for steering the evolution of
highly-excited many-particle system by a strong laser field. The first optimal
control experiment on clusters in strong fields has been conducted by Zamith
et al.~\cite{ZamPRA04}, who used a liquid-crystal-based pulse shaper to
maximize the abundance of highly charged Xe ions.  The optimized light field
contains a two-pulse sequence and produces ions up to Xe$^{23+}$. The spectral
phase of the optimized laser field, however, was not analyzed.

\begin{figure*}[htb]
\centering \resizebox{1.8\columnwidth}{!}{\includegraphics{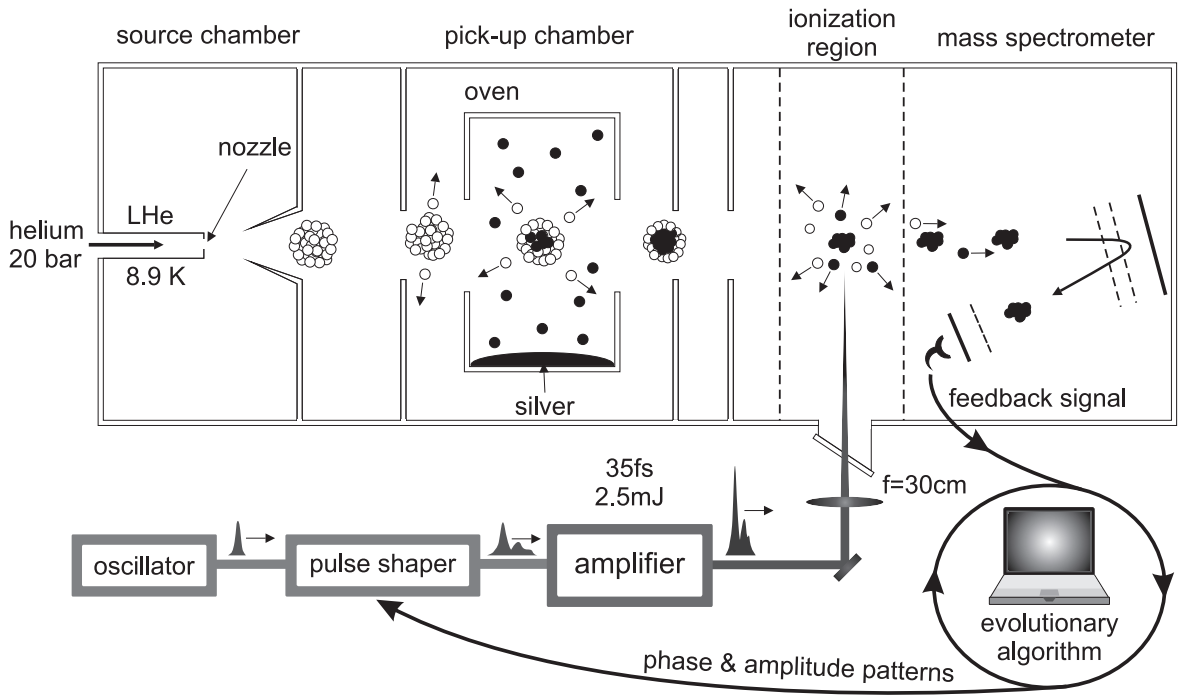}}
\caption{Schematic view of the experiment: The top part visualizes the
  generation of the helium nanodroplets, the formation of the silver clusters
  inside the droplets by sequential atomic pickup in the oven, the laser
  excitation in the interaction region, and the time-of-flight analysis in the
  reflectron time-of-flight mass spectrometer (from left to right).  The
  bottom part sketches the stages of pulse delivery from seed pulse generation
  in the oscillator, over pulse modification in the pulse shaper and final
  amplification, to pulse focusing into the molecular beam apparatus. In the
  optimal control experiment the yield of highly charged ions is maximized by
  applying an evolutionary algorithm scheme within a feedback loop (see
  arrows).}
\label{fig:expschema}
\end{figure*}

In the present contribution we report on an optimal control experiment
including a complete optimization and characterization of amplitude and phase
of the laser field.  As a prototype system we study silver clusters embedded
in helium nanodroplets, which have shown a high pulse-structure sensitivity in
previous experiments~\cite{DoePRL05}. For this metallic system collective
electron motion is expected to dominate the optical absorption. In order to
maximize the abundance of highly charged Ag$^{q+}$ ions a multi-parameter
genetic algorithm procedure is used for tailoring the pulses. A sketch of the
experimental procedure is displayed in Fig.\,\ref{fig:expschema}.

To evaluate the effect of the optimization, the resulting ion yields are
compared to the spectra produced with bandwidth-limited and dispersively
stretched pulses. The optimal pulse shapes are characterized by
frequency-resolved optical gating (FROG)~\cite{Trebino00} and compared to
results from a computational control experiment. The numerical optimization,
based on a modified nanoplasma model~\cite{HilLP09}, varies the pulse envelope
function and is performed for a simplified system, i.e., pure silver
clusters.

The remainder of the text is organized as follows. Section~\ref{sec:methods}
describes the experimental methods, i.e., the cluster generation, the laser
system, and the details of pulse shaping and characterization. The results of
the experimental and numerical closed-loop optimizations are presented in
Sec.~\ref{sec:results}. The discussion and comparison of the data are subject
of Sec.~\ref{sec:discussion}. Finally, conclusions are summarized in
Sec.~\ref{sec:conclu}.

\section{Methods}
\label{sec:methods}
\subsection{Experimental setup}
\label{sec:clustersource}

Silver clusters are produced by the helium droplet pickup
technique~\cite{BarPRL96}, see Fig.\,\ref{fig:expschema}. Superfluid helium
nanodroplets are generated by means of a supersonic expansion of highly
purified helium gas (He\,6.0) at a stagnation pressure of 20\,bars and a
temperature of 8.9\,K through a 5-$\mu$m nozzle yielding a mean droplet size
of $10^{6}$ atoms.  After differential pumping the molecular beam enters the
pickup region where silver atoms are loaded into the droplet, forming
clusters consisting of up to 150 atoms~\cite{DieJCP02}. By changing the nozzle
temperature and the metal vapor pressure in the pickup cell, the size of the
droplet as well as the size of the metal core can be varied nearly
independently over a broad range. We note that this type of cluster source
shows an outstanding long-term stability, which is a central requirement for
performing the optimization experiments.

\begin{figure}[tb]
\resizebox{0.9\columnwidth}{!}{\includegraphics{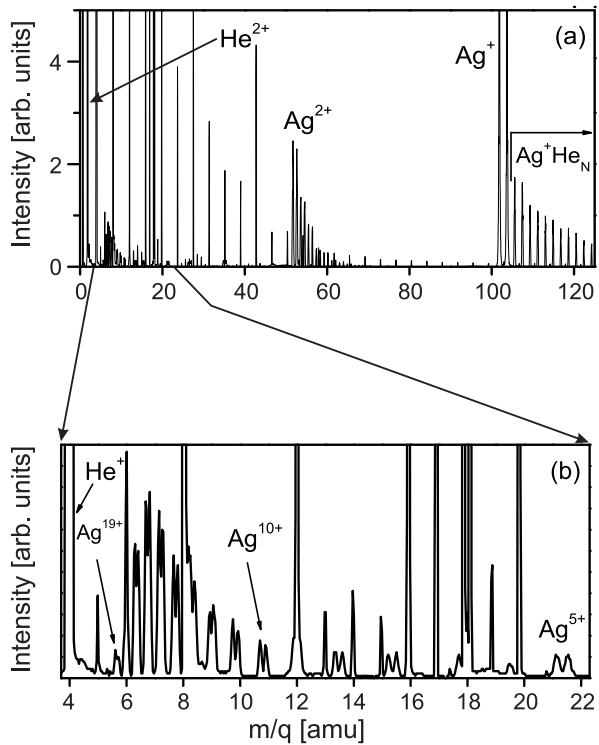}}
\caption{Time-of-flight mass spectrum resulting from excitation of silver
  clusters in helium droplets by stretched 500\,fs laser pulses at intensity
  $7\times10^{14}$ W/cm$^{2}$ (top panel: overview; bottom panel: enlarged
  section highlighting the range from Ag$^{5+}$ to Ag$^{19+}$). The double
  peak structure of individual ion peaks results from the isotopic pattern of
  the silver atom ($^{107}$Ag, $^{109}$Ag). Due to the high recoil energies
  from the cluster Coulomb explosion the Ag ion features are broadened but can
  still be clearly resolved with the chosen detection technique.  In addition
  to Ag$^{q+}$, also He$^+$, He$^{++}$, helium cluster ions, and
  Ag$^{q+}$He$_N$ are observed.}
\label{fig:singleTOF}
\end{figure}

A Ti:Sapphire laser system, consisting of the oscillator, the pulse shaper,
and the pulse amplifier, see Fig.\,\ref{fig:expschema}, delivers pulses with
energies of up to 2.5\,mJ and pulse durations down to 35\,fs (\emph{full width
  half maximum}. FWHM) at 1\,kHz repetition rate. Laser pulses enter the
interaction region perpendicularly to the cluster beam and are focused by a
30\,cm lens (f$^{\#}$\,=\,30) to intensities of up to $10^{16}$\,W/cm$^{2}$.
Ions resulting from the laser-cluster interaction are accelerated by a static
electric field of 2\,kV/\,cm and analyzed by reflectron time-of-flight mass
spectrometry~\cite{DoePCCP07}. Fig.\,\ref{fig:singleTOF} shows a typical mass
spectrum resulting from excitation of the embedded metal complexes with
500\,fs pulses at intensity \mbox{$I_0=7\times10^{14}$\,W/cm$^{2}$}. The
progression of multiply charged Ag$^{q+}$ ions with $q$ up to 19 can clearly
be resolved with the applied detection technique, see
Fig.\,\ref{fig:singleTOF}(b).

\subsection{The Pulse Shaper}
\label{TheDazz}
The oscillator delivers seed pulses with a spectral bandwidth of 50\,nm
(FWHM).  Before amplification the pulses enter an acousto-optic programmable
dispersive filter (AOPDF)(Dazzler, Fastlite) for pulse shaping, see
Fig.\,\ref{fig:expschema}.  The shaper modulates both the amplitude and the
phase of the laser pulses simultaneously according to
\begin{equation}
E_{out}(\omega)\propto S(\omega)E_{in}(\omega),
\end{equation}
with $E_{in}(\omega)$ and $E_{out}(\omega)$ the input and output laser fields
in the frequency domain, and $S(\omega)$ the frequency response function of
the filter.  Arbitrary femtosecond laser pulse shapes can be generated with
the appropriate $S(\omega)$ by controlling the acousto-optical signal. The
frequency response function reads
\begin{equation}
S(\omega)= A(\omega)\exp[{i B (\omega)}],
\end{equation}
where $A(\omega)$ and $B(\omega)$ are real functions describing the spectral
amplitude and phase modulation, respectively. The AOPDF allows to realize a
maximum group delay of 3\,ps with 0.6\,nm spectral resolution. For a more
detailed description see Refs.~\cite{VerOL00,DAZZLER}.  In our setup,
$A(\omega)$ and $B(\omega)$ are defined by 60 parameters (30 in the range $[0,
1]$ for $A(\omega)$ and 30 in the range $[0, 2\pi]$ for $B(\omega)$) equally
spaced over the spectral range.

In the control experiments these parameters are varied to optimize the pulse
structure. We use an extended evolutionary algorithm scheme~\cite{PohBoo00} to
find the optimal laser pulse that maximizes the yield of highly charged atomic
ions. For each pulse shape the mass spectrum is averaged over 6000 laser
shots, whereas 50 individual laser pulse shapes enter each optimization cycle. The
population of different pulse shapes is divided into separate sub--populations
whose constituents are subject to crossover and random mutation after each
generation.  This procedure has been proven to be more robust and efficient
when compared to simple genetic schemes~\cite{Goldberg89}. The evolutionary
process is iterated until the fitness value levels out. A full optimization
experiment roughly takes 3-4 hours. The final result of the optimization
procedure, i.e., the laser pulse structure yielding the highest fitness value, is fully
characterized in amplitude and phase by a home-build frequency-optical-gating
(FROG) analyzer~\cite{Trebino00}.

\section{Results}
\label{sec:results}

\subsection{Experimental findings}
\label{ExpResults}
\begin{figure}[t]
\resizebox{0.9\columnwidth}{!}{\includegraphics{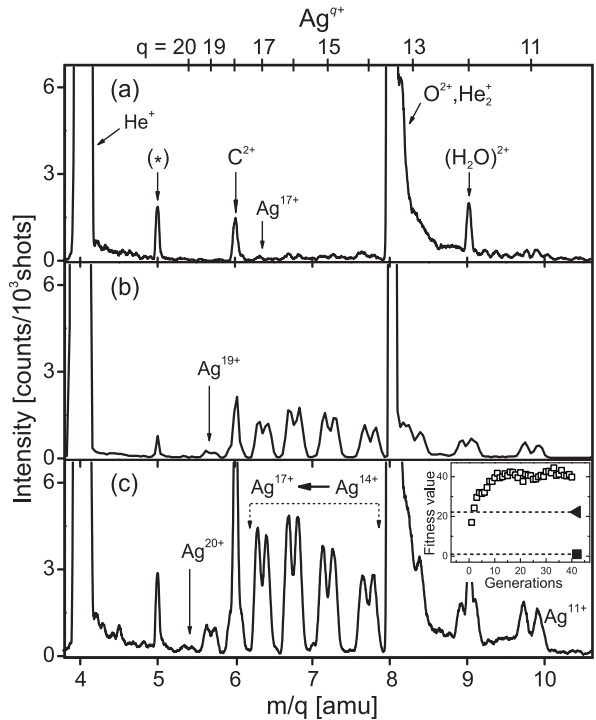}}
\caption{Mass spectra with highly charged atomic ions obtained when exposing
  embedded silver clusters to intense laser pulses of different pulse shapes:
  (a) 50\,fs pulses; (b) 500\,fs linearly down-chirped pulses (cf.
  Fig.\,\ref{fig:singleTOF}); (c) laser pulses fully optimized in amplitude
  and phase within the control experiment.  Note, that the scales are directly
  comparable.  Obviously the yield of Ag$^{q+}$ is substantially enhanced with
  the tailored pulses.  The inset in (c) shows the evolution of the average
  fitness value in the optimization experiment, i.e., the total yield of
  Ag$^{q+}$ (q=14-17) (open squares) compared to the reference measurements
  (50\,fs: full square; 500\,fs pulses: full triangle). }
\label{fig:optispec}
\end{figure}

A comparison of the ion spectra resulting from excitation of the embedded
silver clusters with (i) bandwidth-limited, (ii) optimally stretched, and
(iii) fully shaped laser pulses is given in Fig.\,\ref{fig:optispec}.
The data in Fig.\,\ref{fig:optispec}(a) shows a spectrum for cluster
excitations with 50\,fs pulses (i) at peak intensity $I_0=10^{16}$\,W\,/cm. In
this case the ion signal is dominated by He$_N$ ions and weakly charged
species of residual gas molecules (e.g., C$^{2+}$).  Zooming in, however,
reveals the generation of highly charged silver ions up to Ag$^{17+}$ from the
clusters with low yields, e.g., with about $1.5\times10^{-3}$ ions per laser
shot for Ag$^{17+}$. From previous studies it is well-known that longer pulses
enhance the ion yield, see e.g.~\cite{KoePRL99}.  We utilize the AOPDF to
stretch the pulses (ii) by introducing a linear temporal down--chirp on the
frequency component.  This results in a substantial increase in the yield of
multiply-charged species from the clusters while ion contributions from
background gas decline.  At a pulse width of about 500\,fs
($I_o=10^{15}$\,W/\,cm,$^{-2}$), see Fig.\,\ref{fig:optispec}(b), we observe
the maximal signal of Ag$^{q+}$, indicating optimal ionization conditions with
a single pulse.  Ag$^{19+}$ shows up in the spectrum as the highest charge
state. The two single-pulse measurements (i, ii) serve as references for the
optimization experiments.

For the full optimization procedure (iii) the yield of Ag$^{q+}$ ($q=14-17$)
is taken as fitness value for the genetic algorithm. The resulting final mass
spectrum is given in Fig.\,\ref{fig:optispec}(c). The evolution of the fitness
value (see inset) shows that the pulse structure quickly adapts to the details
of the cluster dynamics. The fitness value exceeds the value for the optimally
stretched single pulse (solid triangle) already after a few iterations.
Subsequently, the fitness value begins to level out and saturates at about
twice the stretched pulse value. Hence, the high-$q$ ion yield with the fully
shaped light field is enhanced by more than a factor of two when compared to
the optimal single pulse, see Fig.\,\ref{fig:optispec}(b) and (c). The
increase is even more pronounced (factor 17) when comparing to the spectra for
the shortest and thus most intense pulse, cf. Fig.\,\ref{fig:optispec}(a). In
addition, we also observe an increase in the highest charge state reaching
$q_{max}=20$ with the fully optimized pulse shape.

\begin{figure}[t]
\resizebox{\columnwidth}{!}{%
  \includegraphics{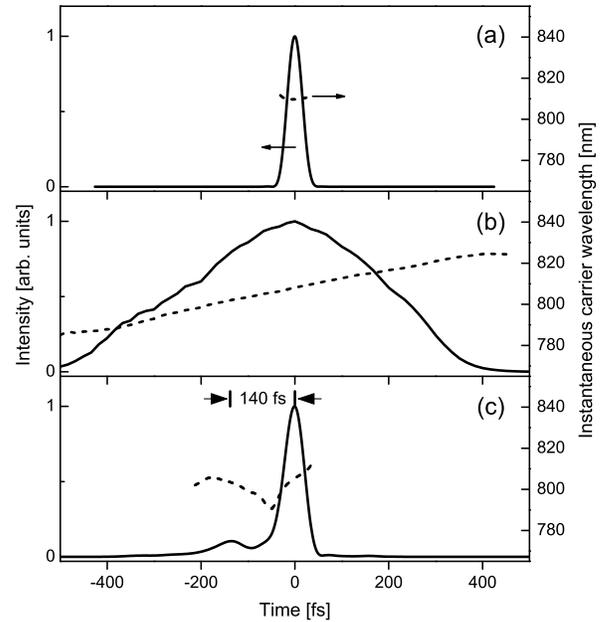} }
\caption{Pulse shape (solid) and evolution of the carrier wavelength (dashed)
  of the laser fields corresponding to the three excitation scenarios in
  Fig.~\ref{fig:optispec}: (a) 50\,fs pulse; (b) 500\,fs linearly down-chirped
  pulse; (c) fully optimized light field.  The tailored laser field leading to
  the enhanced generation of highly charged ions exhibits a double pulse
  structure, comprised of a weak, slowly rising prepulse and a stronger main
  pulse separated by 140\,fs.}
\label{fig:optiPulse}
\end{figure}
As a first step towards unraveling the underlying dynamics, the pulses from
the three excitation scenarios in Fig.\,\ref{fig:optispec}(a)-(c) have been
characterized in amplitude and phase, see Fig.\,\ref{fig:optiPulse}(a)-(c).
For convenience, the temporal intensity is given (solid) together with the
phase evolution in terms of the instantaneous carrier wavelength (dashed).
which is derived from the instantaneous angular frequency
$\omega(t)=\omega_0-\dot{\phi}(t)$, with $\omega_0$ being the central angular
frequency and $\phi(t)$ the temporal phase of the pulse \cite{Trebino00}.

Whereas the flat evolution of the instantaneous carrier wavelength in
Fig.\,\ref{fig:optiPulse}(a) underlines excitation with a nearly
bandwidth-limited pulse, the linear wavelength shift in
Fig.\,\ref{fig:optiPulse}(b) reflects the dispersive pulse stretching. The
analysis of the optimized laser field, see Fig.~\ref{fig:optiPulse}(c), shows
a simple double-pulse intensity distribution (solid line). A prepulse
containing about 15\% of the pulse energy excites the clusters prior to the
main pulse.  The pulse intensity drops significantly between the subpulses,
indicating a two--step excitation process to induce the effective generation
of high charge states. This signature of the pulse is found to be rather
robust with respect to the chosen pulse energy, i.e., the pulse delay as well
as the intensity ratio may change while the general shape is preserved. Note
that the optimization procedure introduces a substantial red-shift of the
instantaneous carrier wavelength from 790\,nm to 810\,nm during the main
pulse, see dashed curve in Fig.~\ref{fig:optiPulse}(c).

\subsection{Computational results}
\label{CompResults}
To assist the experimental study, we apply a theoretical approach to find
laser pulse shapes that maximize the yield of highly charged ions in the
laser-cluster interaction. For this purpose, a genetic algorithm is used and
adapted to the nanoplasma approach of Ditmire~\cite{DitPRA96}. The model
contains the essential physical processes like ionization, heating and
expansion. Briefly, the clusters are assumed to have uniform temperature and
density profiles and are initialized as neutral spheres. To model the nanoplasma inner ionization dynamics, tunneling as well as
electron impact ionization are taken into account. Heating via inverse
bremsstrahlung, being the dominant process for laser energy capture, is
described with effective rates for collisional absorption. The cluster
expansion and the resulting cooling of the system are calculated using
hydrodynamic equations. In our version of the nanoplasma model, plasma heating
is treated by a generalized quantum statistical expression for electron-ion
collisions including resonant collective absorption ~\cite{BorCPP07}. In addition, damping due to electron collisions with the cluster boundary, i.e., surface friction, is incorporated by including a corresponding collision term \cite{MegJPB03,MicPRA08}. Furthermore, the lowering of the ionization energies due to screening is taken into account~\cite{HilLP09}.

To calculate the optimal light field for producing a maximum yield of specific
ion species a genetic feedback algorithm is applied. For the optimization
procedure, a pulse intensity shape $I(t)$ according to the Ansatz
\begin{equation}
I(t)=\sum_{n=0}^4 A_n \exp \left[ -\frac{4\,{\rm ln}\,2\,(t-t_n)^2}{\sigma_n^2}\right]
\end{equation}
\noindent
is used, with $A_n$, $t_n$, and $\sigma_n$ being the weights, time offsets,
and temporal width parameters (FWHM) of the individual terms. Note that a
fixed carrier wavelength of 810\,nm is used in the calculations, i.e., the
phase evolution is not optimized. The total pulse fluence is kept constant for
each chosen pulse--shape and corresponds to a 810\,nm Gaussian pulse of 35\,fs
(FWHM) with peak intensity $1\times10^{16}$\,W/cm$^{2}$. Further, a lower
limit of 35\,fs is used for the width parameters $\sigma_n$. To calculate the
best pulse shape, 20 different pools, each with a population of 40 species,
are used in the optimization procedure.

\begin{center}
\begin{figure}[t]
\resizebox{0.9\columnwidth}{!}{\includegraphics{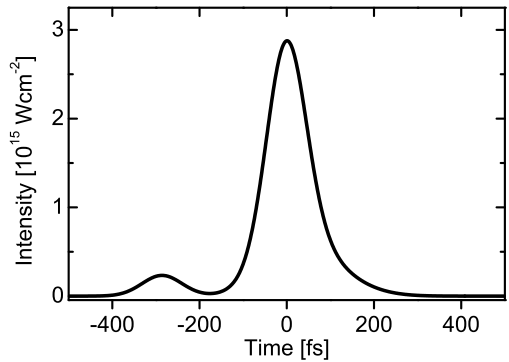} }
\caption{Calculated temporal intensity evolution of the optimal light field
  for producing a maximum yield of Ag$^{10+}$. The intensity envelope is in
  qualitative accordance with the experimental result, cf.
  Fig.~\ref{fig:optiPulse}. The total pulse fluence corresponds to a 35\,fs
  pulse with peak intensity $1\times10^{16}$\,W/cm$^{2}$.  }
\label{Fig:optipulse_numeric}
\end{figure}
\end{center}

We perform the calculations as close as possible to the experimental
conditions. Silver clusters with an initial radius of 2.5\,nm (4500 atoms) are
considered since smaller particles cannot be treated adequately within the
nanoplasma model. The calculated optimal pulse shape giving the maximum yield
of Ag$^{10+}$ is shown in Fig.~\ref{Fig:optipulse_numeric}. This ion species
was chosen because higher ionization levels were only weakly populated. Here,
further improvements of the model would be necessary. For a discussion of additional processes to be included, like the enhancement
of electron impact ionization by local cluster fields and the influence of the
experimental setup, especially the frustration of recombination by the ion
extraction electric fields after the laser pulse, see Ref.~\cite{FenPRL07b}. 

As references to assess the results of the genetic algorithm, calculations were also performed for a single 500\,fs pulse and for a dual-pulse excitation scheme. Whereas the single pulse mainly produces Ag$^{7+}$ and Ag$^{8+}$, Ag$^{8+}$ and Ag$^{9+}$ are the dominant contributions in the ion spectra for dual pulse excitation. With the light field from the computational control experiment, see Fig.~\ref{Fig:optipulse_numeric}, the Ag$^{10+}$ yield is enhanced by a factor of four over the result for dual-pulse excitation and by more than a factor of eight over the result for the 500\,fs single pulse.
The simulations predict an optimized light field having a low-intensity prepulse followed by a more intense main pulse, in good agreement with the experimental control study.  Even the intensity ratio between the two subpulses is nearly reproduced by the calculation.

\section{Discussion}
\label{sec:discussion}
The applied pulse shaping technique combined with a genetic algorithm has been
shown to allow the enhanced production of highly charged ions from
strong-field laser-cluster interactions. Compared to simpler excitation
schemes with bandwidth-limited or stretched pulses, the ion yields could be
substantially increased. The laser field tailored in amplitude and phase thus
offers an alternative and more efficient ionization pathway.

We find that a sequence of a weak prepulse followed by a much stronger main
pulse after about 140\,fs leads to the highest ionization under the
experimental boundary conditions. Interestingly, the bandwidth-limited pulse
from the reference scenario and the main feature of the optimized light field
are similar in intensity and duration. This underlines the decisive effect of
the low-intensity prepulse ($10^{15}$W/\,cm$^2$) for enhancing the
ionization efficiency by up to an order of magnitude. The optimized light
field also outperforms optimally stretched single pulses (500\,fs) by a factor
of two in the ion yield.

The envelope of the tailored light field and comparison with the results for a
bandwidth-limited pulse suggest that efficient energy transfer and strong
ionization take place within the stronger main pulse. These findings can be
related to previous dual pulse studies on Ag$_N$ in He
droplets~\cite{DoePRL05}, where a specific optical delay was found to maximize
the highly charged ion yield for excitation with a pair of identical pulses.
The optimization presented here extends these experiments. The control study
not only underlines the importance of a certain pulse delay that matches the
cluster expansion required for most efficient resonant collective heating, but
also highlights the impact of the pulse intensity ratio.  The genetic
optimization shows that a less intense prepulse is favorable for preionization
of the particle in order to launch the cluster expansion. Near resonance, the
pulse adapts to maximum heating and strongest charging. This general scenario
is corroborated by the fact, that the observed characteristic pulse structure
is quite robust with respect to changes of the pulse energy.

The observed pulse shape shows similarities with the result of an earlier
control experiment by Zamith et al.~\cite{ZamPRA04} on Xe$_N$
($\overline{N}\approx1.6\times10^4$). The authors also found an asymmetric
dual pulse with a time separation of 146\,fs. Although the pulse envelope
indicates a similar scenario, the actual target conditions are quite
different. In silver clusters the delocalized valence electrons already form a
metallic state, whereas a rare-gas system requires preionization for the
nanoplasma formation.

Comparing the outcome of the present experimental optimization with the
numerical one shows similar pulse structures. A detailed analysis of the
simulation run for the numerically optimized light field corroborates that the
main energy transfer into the system takes place upon the passage of the
plasmon resonance in the trailing main pulse. Here, the major fraction of the
energy capture occurs leading to strong heating and finally to the formation
of a large number of highly charged ions.  The computational results thus
support that the optimal light field "tries" to utilize most efficient plasmon
enhancement ionization~\cite{KoePRL99}.

Interestingly, the analysis of the instantaneous phase reveals a red-shift of
the central carrier wavelength (down-chirp) during the main laser pulse, cf.
Fig.~\ref{fig:optiPulse}(c).  This behavior can be explained with dynamic
frequency locking of the laser field to the plasmon.  As the cluster expands,
a down-chirped laser field allows extended frequency matching with the
collective mode. For this to occur, however, the increase in the ion density
due to cluster inner ionization has to be overcompensated by a sufficiently
rapid cluster expansion.

Finally, we like to discuss the impact of the helium nanomatrix. Whereas
helium is almost transparent at infrared wavelengths for low intensities,
substantial ionization and heating may occur under the strong laser field. In
our calculations the helium is not taken into account up to now due to the
restricted dimensional approach of the nanoplasma model.  Very likely, the
surrounding helium droplet influences both the timing as well as the final
ionization state of the cluster ions. Further improvements of the model for
resolving the radial structure of the system are underway. As has been
reported recently by Mikaberidze et al.~\cite{MikPRA08} for embedded xenon
clusters (Xe$_{100}$He$_{1000}$), the presence of the helium nanomatrix
results in a large contribution to the total energy absorption in a strong
laser field.  The impact of the helium matrix on the final charge state
distribution of the cluster constituents as probed in our experiment, however,
remains a challenging problem for future investigations.

\section{Conclusion}\label{sec:conclu}
Amplitude-- and phase--modulated pulses have been used to maximize the highly
charged ion yield from silver clusters embedded in helium nanodroplets. A
substantial increase is found when comparing to bandwidth-limited and
optimally stretched pulses. The laser field resulting from the control
experiment shows a two-pulse structure which is in fair agreement with the
result of a computational optimization experiment based on the nanoplasma
model. The shape and phase evolution of the optimized laser field indicate
that the tailoring adapts to take advantage of the plasmon enhanced response.
This proof of principle experiment on the metallic prototype system thus
reveals a unique signature from collective electron excitations and their
importance for controlling strong-field laser-matter interactions.

\begin{acknowledgments}
  Main parts of the helium droplet machine have been provided by J.P. Toennies
  and his group at the MPI G\"ottingen. Financial support by the Deutsche
  Forschungsgemeinschaft within the
  Sonderforschungsbereich SFB 652 is gratefully acknowledged.\\
\end{acknowledgments}

\bibliographystyle{unsrt}

\end{document}